\documentclass[a4paper,twoside]{article}
\usepackage[INRIA]{lipinria}

\usepackage{paralist}

\usepackage[latin1]{inputenc}
\usepackage[T1]{fontenc}
\usepackage{amsthm, amsmath,amsfonts, amssymb, latexsym, amscd}
\usepackage{figlatex}
\usepackage{xcolor}
\usepackage{xspace}
\usepackage{rotating}
\usepackage{url}

\newcommand{\Ss}[1]{\ensuremath{Comm^{start}_{#1}}}
\newcommand{\Se}[1]{\ensuremath{Comm^{end}_{#1}}}
\newcommand{\Cs}[1]{\ensuremath{Comp^{start}_{#1}}}
\newcommand{\Ce}[1]{\ensuremath{Comp^{end}_{#1}}}
\newcommand{\fr}[3]{\ensuremath{\gamma^{#2}_{#1}(#3)}}

\newtheorem{theorem}{Theorem}

\newcommand{\singleinst}{\textsc{SingleInst}\@\xspace}
\newcommand{\multiinst}{\textsc{MultiInst}\@\xspace}
\newcommand{\naive}{\textsc{Simple}\@\xspace}
\newcommand{\LProg}[1]{\textsc{LP}\@\xspace\ensuremath{#1}\@\xspace}
\newcommand{\heurB}{\textsc{Heuristic B}\@\xspace}
\newcommand{\singleload}{\textsc{SingleLoad}\@\xspace}

\usepackage{RR}
\begin{document}

\RRItheme{\THNum}

\RRIprojet{GRAAL}

\RRInumber{2007-29}
\RRNo{6235}

\RRItitle{Scheduling multiple divisible loads\\ on a linear processor network}
\RRIthead{Scheduling multiple divisible loads on a linear processor network}

\RRItitre{Ordonnancement de tâches divisibles sur un réseau linéaire de processeurs}

\RRIauthor{Matthieu Gallet\and Yves Robert\and Frédéric Vivien}
\RRIahead{M. Gallet\and Y. Robert\and F. Vivien}

\RRIdate{June 2007}

\RRIkeywords{scheduling, heterogeneous processors, divisible loads, single-installment, multiple-installments.}
\RRImotscles{ordonnancement, ressources hétérogènes, tâches divisibles, tournées.}

\RRIabstract{
  Min, Veeravalli, and Barlas have recently proposed strategies to
  minimize the overall execution time of one or several divisible
  loads on a heterogeneous linear network, using one or more
  installments~\cite{WongVe04,WongVeBa05}. We show on a very simple
  example that their approach does not always produce a solution and
  that, when it does, the solution is often suboptimal.  We also show
  how to find an optimal schedule for any instance, once the number of
  installments per load is given. Then, we formally state that any
  optimal schedule has an infinite number of installments under a
  linear cost model as the one assumed in~\cite{WongVe04,WongVeBa05}.
  Therefore, such a cost model cannot be used to design practical
  multi-installment strategies. Finally, through extensive simulations
  we confirmed that the best solution is always produced by the linear
  programming approach, while solutions of~\cite{WongVeBa05} can be
  far away from the optimal.
}

\RRIresume{Min, Veeravalli, and Barlas ont
  proposé~\cite{WongVe04,WongVeBa05} des stratégies pour minimiser le
  temps d'exécution d'une ou de plusieurs tâches divisibles sur un
  réseau linéaire de processeurs hétérogènes, en distribuant le
  travail en une ou plusieurs tournées. Sur un exemple très simple
  nous montrons que l'approche proposée dans~\cite{WongVeBa05} ne
  produit pas toujours une solution et que, quand elle le fait, la
  solution est souvent sous-optimale. Nous montrons également comment
  trouver un ordonnancement optimal pour toute instance, quand le
  nombre de tournées par tâches est spécifié. Ensuuite, nous montrons
  formellement que lorsque les fonctions de coûts sont linéaires,
  comme c'est le cas dans~\cite{WongVe04,WongVeBa05}, un
  ordonnancement optimal a un nombre infini de tournées. Un tel modèle
  de coût ne peut donc pas être utilisé pour définir des stratégies en
  multi-tournées utilisables en pratique. Finalement, au moyen de
  simulations exhaustives, nous montrons que la meilleure solution est
  toujours produite par l'approche par programmation linéaire, tandis
  que les solutions de~\cite{WongVeBa05} peuvent être très éloignées
  de l'optimal.}

\RRImaketitle

\section{Introduction}
\label{sec.intro}

Efficiently scheduling the tasks of a parallel application onto the
resources of a distributed computing platform is critical for
achieving high performance. This scheduling problem has been studied
for a variety of application models. Some popular models consider a
set of independent tasks without task synchronization nor
inter-task communications. Among these models some focus on the case
in which the number of tasks and the task sizes can be chosen
arbitrarily. This corresponds to the case when the application
consists of an amount of computation, or \emph{load}, that can be
arbitrarily divided into any number of independent pieces of arbitrary
sizes. This corresponds to a perfectly parallel job: any sub-task can
itself be processed in parallel, and on any number of workers. In
practice, this model is an approximation of an application that
consists of (very) large numbers of identical, low-granularity
computations. This \emph{divisible load} model has been widely studied
in the last several years, and \emph{Divisible Load Theory} (DLT) has
been popularized by the landmark book written in 1996 by Bharadwaj,
Ghose, Mani and Robertazzi~\cite{robertazzi96}.  DLT has been applied
to a large spectrum of scientific problems, including linear
algebra~\cite{ChanBG01}, image processing~\cite{LeeHam95,LiBK03},
video and multimedia broadcasting~\cite{AltilarPak98,AltilarPak02},
database searching~\cite{BlazewiczDM99}, biological
pattern-matching~\cite{LegrandSuVi06}, and the processing of large
distributed files~\cite{WangKMAC98}.

Divisible load theory provides a practical framework for the mapping
of independent tasks onto heterogeneous platforms. From a theoretical
standpoint, the success of the divisible load model is mostly due to
its analytical tractability. Optimal algorithms and closed-form
formulas exist for the simplest instances of the divisible load
problem. We are aware of only one NP-completeness result in the
DLT~\cite{LegrandNP}. This is in sharp contrast with the theory of
task graph scheduling, which abounds in NP-completeness theorems and
in inapproximability results.

Several papers in the Divisible Load Theory field consider
master-worker platforms~\cite{robertazzi96,GenaudGiVi2004,j89}. 
However, in
communication-bound situations, a linear network of processors can
lead to better performance:  on such a platform, several
communications can take place simultaneously, thereby  enabling a pipelined
approach. Recently, Min, Veeravalli, and Barlas have
proposed strategies to minimize the overall
execution time of one or several divisible loads on a heterogeneous
linear processor network~\cite{WongVe04,WongVeBa05}. Initially, the authors targeted single-installment
strategies, that is strategies under which a processor receives in a
single communication all its share of a load. But for cases where their approach failed
to design single-installment strategies, they also considered
multi-installment solutions.

In this paper, we first show on a very simple example
(Section~\ref{sec.example}) that the approach proposed
in~\cite{WongVeBa05} does not always produce a solution and that, when
it does, the solution is often suboptimal.  The fundamental flaw of
the approach of~\cite{WongVeBa05} is that the authors are optimizing
the scheduling load by load, instead of attempting a global
optimization. The load by load approach is suboptimal and
unduly over-constrains the problem.
On the contrary, we show (Section~\ref{sec.new}) how to find an
optimal scheduling for any instance, once the number of installments
per load is given. In particular, our approach always find the optimal
solution in the single-installment case. We also formally state
(Section~\ref{sec:ext}) that under a linear cost model for
communication and communication, as in~\cite{WongVe04,WongVeBa05}, an
optimal schedule has an infinite number of installments. Such a cost
model can therefore not be used to design practical multi-installment
strategies. Finally, in Section~\ref{sec.experiments}, we report the
simulations that we performed in order to assess the actual efficiency of the
different approaches. We now start by introducing the framework.

\section{Problem and Notations}
\label{sec.frame}

\begin{figure}
\begin{center}
	{\includegraphics[width=0.75\linewidth]{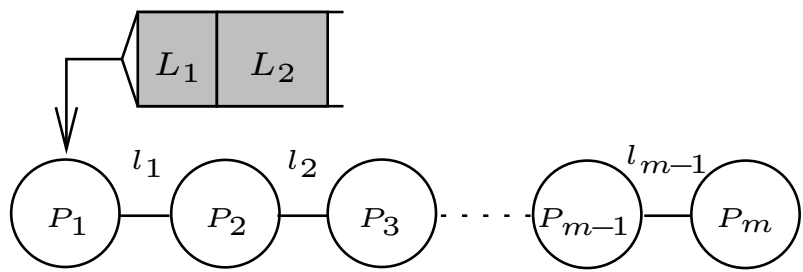}}
        \caption{Linear network, with $m$ processors and $m-1$ links.}
        \label{Fig:platform}
\end{center}
\end{figure}

We use a framework similar to that of~\cite{WongVe04,WongVeBa05}.  The
target architecture is a linear chain of $m$ processors ($P_{1}, P_2,
\ldots, P_{m}$).  Processor $P_{i}$ is available from time $\tau_i$.
It is connected to processor $P_{i+1}$ by the communication link
$l_{i}$ (see Figure~\ref{Fig:platform}).  The target application is
composed of $N$ loads, which are \emph{divisible}, which means that
each load can be split into an arbitrary number of chunks of any size,
and these chunks can be processed independently.  All the loads are
initially available on processor $P_{1}$, which processes a fraction
of them and delegates (sends) the remaining fraction to $P_2$.  In
turn, $P_2$ executes part of the load that it receives from $P_1$ and
sends the rest to $P_3$, and so on along the processor chain.
Communications can be overlapped with (independent) computations, but
a given processor can be active in at most a single communication at
any time-step: sends and receives are serialized (this is the full
\emph{one-port} model).

Since the last processor $P_m$ cannot start computing before having
received its first message, it is useful for $P_1$ to distribute the
loads in several installments: the idle time of remote processors in
the chain will be reduced due to the fact that communications are
smaller in the first steps of the overall execution.

The objective is to minimize the \emph{makespan}, i.e., the time at
which all loads are completed. For the sake of convenience, all
notations are summarized in Table~\ref{notations}.

We deal with the general case in which the $n$th load is distributed
in $Q_{n}$ installments of different sizes. For the $j$th installment
of load $n$, processor $P_i$ takes a fraction \fr{j}{n}{i}, and sends
the remaining part to the next processor while processing its own
fraction.

Loads have different characteristics: load $n$ (with $1 \leq n \leq
N$) is defined by a volume of data $V_{comm}(n)$ and a quantity of
computation $V_{comp}(n)$. Moreover, processors and links are not
identical either. We let $w_{i}$ be the time taken by $P_{i}$ to
compute a unit load ($1\leq i\leq m$), and $z_{i}$ be the time taken
by $P_{i}$ to send a unit load to $P_{i+1}$ (over link $l_i$, $1\leq i
\leq m-1$).  Note that we assume a linear model for computations and
communications, as in the original
articles~\cite{WongVe04,WongVeBa05}, and as is often the case in
divisible load
literature~\cite{Robertazzi-Computer2003,cluster-special} (we will
discuss this hypothesis in Section~\ref{sec:ext}).

For the $j$th installment of the $n$th load, let $\Ss{i,n,j}$ denote
the starting time of the communication between $P_{i}$ and $P_{i+1}$,
and let $\Se{i,n,j}$ denote its completion time; similarly,
$\Cs{i,n,j}$ denotes the start time of the computation on $P_{i}$ for
this installment, and $\Ce{i,n,j}$ denotes its completion time.
Following~\cite{WongVe04,WongVeBa05}, we make the assumption that
processor $P_i$ sends the relevant fraction of the $j$th installment
of the $n$th load to processor $P_{i+1}$ \emph{before} it starts to
receive another fraction of load from $P_{i-1}$. 
Similarly, we suppose that the order in which
the different application loads are sent is fixed. Although very natural,
these assumptions do reduce the solution space, and it might be useful
to relax them in some special cases.

\begin{table*}
\begin{center}
\begin{tabular}{|ll|}
\hline
$m$ & Number of processors in the system.\\
\hline
$P_{i}$ & Processor $i$, where $i = 1, \ldots, m$.\\
\hline
$w_{i}$ & Time taken by processor $P_{i}$ to compute a unit load.\\
\hline
$z_{i}$ & Time taken by $P_{i}$ to transmit a unit load to $P_{i+1}$.\\
\hline
$\tau_{i}$ & Availability date of $P_{i}$ (time at which it becomes available for processing the loads).\\
\hline
$N$ & Total number of loads to process in the system.\\
\hline
$Q_{n}$ & Total number of installments for $n$th load.\\
\hline
$V_{comm}(n)$ & Volume of data for $n$th load.\\
\hline
$V_{comp}(n)$ & Volume of computation for $n$th load.\\
\hline
$\fr{i}{j}{n}$ & Fraction of $n$th load computed on processor $P_{i}$ during the $j$th installment.\\
\hline
$\Ss{i,n,j}$ & Start time of communication from processor $P_{i}$ to processor $P_{i+1}$\\
& for $j$th installment of $n$th load.\\
\hline
$\Se{i,n,j}$ & End time of communication from processor $P_{i}$ to processor $P_{i+1}$\\
& for $j$th installment of $n$th load.\\
\hline
$\Cs{i,n,j}$ & Start time of computation on processor $P_{i}$\\
& for $j$th installment of $n$th load.\\
\hline
$\Ce{i,n,j}$ & End time of computation on processor $P_{i}$\\
& for $j$th installment of $n$th load.\\
\hline
\end{tabular}
\caption{Summary of notations.\label{notations}}
\end{center}
\end{table*}

\section{Motivating example}
\label{sec.example}

We first recall the algorithms presented in~\cite{WongVeBa05}. We then
introduce our motivating example and use it to assess the performance
of these algorithms.

\subsection{The existing algorithms}
\label{sec.existingalgos}

It is often \emph{stated} that, when scheduling a single load under
the divisible load model, in an optimal solution ``all participating
processors stop computing at the same time
instant''~\cite{WongVeBa05}. This property has been formally proved
for some particular settings~\cite{j89,GenaudGiVi2004} but is used far
more generally and some existing proofs are even flawed
(see~\cite{GenaudGiVi2004} for examples).

Min, Veeravalli, and Barlas use this \emph{optimality principle} to
build their algorithm. They assume that all processors participate in
the processing of each load and all complete simultaneously the
processing of any given load. The strict application of this principle
leads to what we call the \singleinst algorithm. In order to further
optimize the processing of the loads, they force each processor to
stay busy from the first time it starts processing a load to the
overall completion. When such a solution does not exist with a
single-installment strategy, that is when a processor receives in a
single communication all its share of a given load, they resort to
multi-installment strategies where each installment is the largest
possible satisfying all the constraints (all processors complete
simultaneously an installment processing). This defines their main
algorithm, that we call \multiinst. The idea is to fully overlap
communications by computations (which is obviously not always possible
when communications are far more expensive than computations). Both
algorithms optimize the schedule load by load, instead of attempting a
global optimization.

\subsection{The example}
\label{sec.ours}

Our motivating example uses $2$ identical processors $P_{1}$ and
$P_{2}$ with $w_{1} = w_{2} = \lambda$, and $z_1 = 1$.  We consider
$N= 2$ identical divisible loads to process, with
$V_{comm}(1)=V_{comm}(2)=1$ and $V_{comp}(1)=V_{comp}(2)=1$. Note that
when $\lambda$ is large, communications become negligible and each
processor is expected to process around half of both loads. But when
$\lambda$ is close to $0$, communications are very important, and the
solution is not obvious. As both processors have the same
computational power, under \multiinst they will process the same
fraction of any given installment of any given load, except for the
first installment of the first load.

To ease the reading, we only give a short (intuitive) description of
the schedules, and we provide the different makespans without
justification; all details can be found in the research
report~\cite{GalletRoVi07a}.

\begin{figure}
\begin{center}
	{\includegraphics[width=0.75\linewidth]{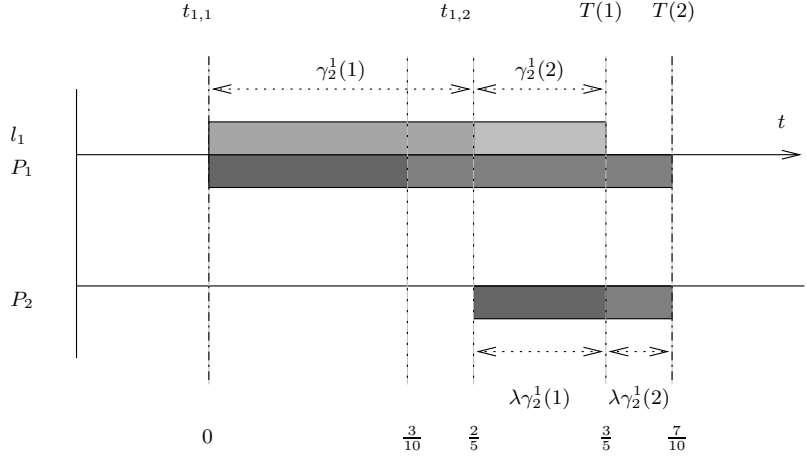}}
        \caption{A possible schedule when $\lambda=\frac{1}{2}$.\label{Fig:example}}
\end{center}
\end{figure}

We first consider a simple schedule which uses a single installment for each load,
as illustrated in Figure~\ref{Fig:example}.
Processor $P_{1}$ computes a fraction $\fr{1}{1}{1} =	\frac{2\lambda^{2}+1}{2\lambda^{2}+2\lambda+1}$ of the first load, and a fraction $\fr{1}{1}{2}	= \frac{2\lambda+1}{2\lambda^{2}+2\lambda+1}$ of the second load.
Then the second processor computes a fraction $\fr{2}{1}{1} = \frac{2\lambda}{2\lambda^{2}+2\lambda+1}$ of the first load, and a fraction $\fr{2}{1}{2}	= \frac{2\lambda^{2}}{2\lambda^{2}+2\lambda+1}$ of the second load.
The makespan achieved by this schedule is equal to $\mathrm{makespan}_{1}=\frac{2\lambda\left(\lambda^{2}+\lambda+1\right)}{2\lambda^{2}+2\lambda+1}$.

\subsection{Case $\lambda\geq \frac{\sqrt{3}+1}{2}$: single-installment}
\label{oneinstallment}

Under the algorithms of~\cite{WongVeBa05}, $P_{1}$ and $P_{2}$ have to
simultaneously complete the processing of their share of the first
load. The same holds true for the second load. We are in the
one-installment case when $P_1$ is fast enough to send the second load
to $P_{2}$ while it is computing the first load (hence \singleinst and
\multiinst have the same output). This condition writes $\lambda\geq
\frac{\sqrt{3}+1}{2}\approx 1.366$. Then, $P_{1}$ processes a fraction
$\fr{1}{1}{1}=\frac{\lambda+1}{2\lambda+1}$ of the first load, and a
fraction $\fr{1}{1}{2}=\frac{1}{2}$ of the second one.
The makespan achieved by this schedule is $\mathrm{makespan}_{2} =
\frac{\lambda\left(4\lambda+3\right)}{2\left(2\lambda+1\right)}$.

\begin{figure}
  \begin{center}
    {\includegraphics[width=0.75\linewidth]{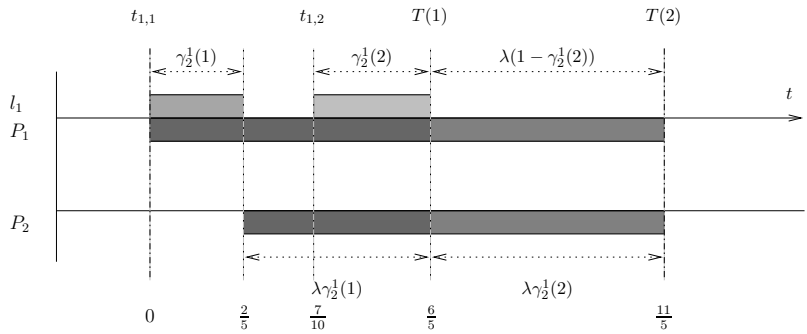}}
    \caption{The schedule of~\cite{WongVeBa05} for $\lambda=2$.\label{Fig:example_l2}}
  \end{center}
\end{figure}

Comparing both makespans, we have $0\leq \mathrm{makespan}_{2} -
\mathrm{makespan}_{1}\leq \frac{1}{4}$, the solution
of~\cite{WongVeBa05} having a strictly larger makespan, except when
$\lambda = \frac{\sqrt{3}+1}{2}$. A visual representation of this case
is given in Figure~\ref{Fig:example_l2} for $\lambda=2$.

\subsection{Case $\lambda < \frac{\sqrt{3}+1}{2}$: multi-installment}
\label{multiinstallment}

The solution of~\cite{WongVeBa05} is a multi-installment strategy when
$\lambda < \frac{\sqrt{3}+1}{2}$, i.e., when communications tend to be
important compared to computations.  More precisely, this case happens
when $P_{1}$ does not have enough time to completely send the second
load to $P_{2}$ before the end of the computation of the first load on
both processors.

The way to proceed in~\cite{WongVeBa05} is to send the second load
using a multi-installment strategy. $Q_2$ denote the number of
installments for this second load.  We can easily compute the size of
each fraction distributed to $P_{1}$ and $P_{2}$.  Processor $P_{1}$
has to process a fraction $\fr{1}{1}{1}=\frac{\lambda+1}{2\lambda+1}$
of the first load, and fractions $\fr{1}{1}{2}, \fr{1}{2}{2},\ldots,
\fr{1}{Q_2}{2}$ of the second one. Moreover, for $1\leq k < Q_2$, due to
all the assumptions, we have $\fr{1}{k}{2}=\lambda^{k}\fr{2}{1}{1}$.
And for $k=Q_2$ (the last installment), we have $\fr{1}{Q_2}{2} \leq
\lambda^{Q_2}\fr{2}{1}{1}$. We can then establish an upper bound on the
portion of the second load distributed in $Q_2$ installments:
$$
\sum_{k=1}^{Q_2}\left(2\gamma^k_{1}(2)\right) \leq
2\sum_{k=1}^{Q_2}\left(\fr{2}{1}{1}\lambda^{k}\right)
=\frac{2\left(\lambda^{Q_2}-1\right)\lambda^{2}}{2\lambda^{2}-\lambda-1}
$$
if $\lambda\neq 1$, and $Q_2=2$ otherwise. We have three cases to
discuss:
\begin{enumerate}
\item {$0 < \lambda < \frac{\sqrt{17}+1}{8} \approx 0.64$:} Since
  $\lambda < 1$, we can write for any nonnegative integer $Q_2$:
  $$\sum_{k=1}^{Q_2}\left(2\gamma_1^k(2)\right)< \sum_{k=1}^{\infty}\left(2\gamma^1_2(1)\lambda^k\right)=\frac{2\lambda^{2}}{(1-\lambda)(2\lambda+1)}$$
  $\frac{2\lambda^{2}}{(1-\lambda)(2\lambda+1)} < 1$ when $\lambda <
  \frac{\sqrt{17}+1}{8}$.  So, an infinite number of installments do
  not suffice to completely process the second load. In other words,
  no solution is found in~\cite{WongVeBa05} for this case. A visual
  representation of this case is given in Figure~\ref{Fig:example_l05}
  with $\lambda=0.5$.
  \begin{figure}
    \begin{center}
      {\includegraphics[width=0.75\linewidth]{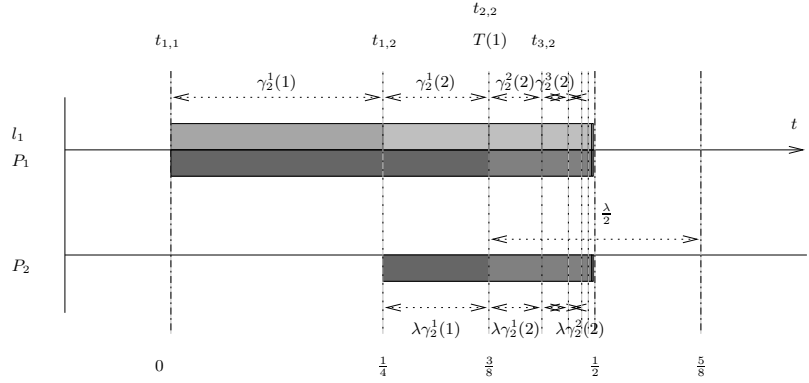}}
      \caption{The schedule of~\cite{WongVeBa05} for $\lambda=\frac{1}{2}$.\label{Fig:example_l05}}
    \end{center}
  \end{figure}

\item{$\lambda = \frac{\sqrt{17}+1}{8}$:} Then
  $\frac{2\lambda^{2}}{(1-\lambda)(2\lambda+1)} = 1$, and an infinite
  number of installments is required to completely process the second
  load. This solution is unrealistic.

\item{$\frac{\sqrt{17}+1}{8} < \lambda < \frac{\sqrt{3}+1}{2}$:} The
  solution of~\cite{WongVeBa05} is then a multi-installment solution
  which is better than any solution using a single installment per
  load. (A visual representation of this case is given in
  Figure~\ref{Fig:example_l1} with $\lambda=1$.) However this solution
  may require a very large number of installments.  Furthermore, this
  solution is not optimal. Indeed, consider the case
  $\lambda=\frac{3}{4}$. The algorithm of~\cite{WongVeBa05} achieves a
  makespan equal to
  $\left(1-\fr{2}{1}{1}\right)\lambda+\frac{\lambda}{2}=\frac{9}{10}$.
  The first load is sent in one installment and the second one is sent
  in $3$ installments, as the number of installments is set
  in~\cite{WongVeBa05} as $Q_2 = \left\lceil
    \frac{\ln(\frac{4\lambda^{2}-\lambda-1} {2\lambda^{2}})}
    {\ln(\lambda)} \right\rceil$.  However, we can come up with a
  better schedule by splitting both loads into two installments, and
  distributing them as follows:
  \begin{compactitem}
  \item Load 1, first round: $P_{1}$ processes $0$ unit;
  \item Load 1, first round: $P_{2}$ processes $\frac{192}{653}$ unit;
  \item Load 1, second round: $P_{1}$ processes $\frac{317}{653}$ unit;
  \item Load 1, second round: $P_{2}$ processes $\frac{144}{653}$ unit;
  \item Load 2, first round: $P_{1}$ processes $0$ unit;
  \item Load 2, first round: $P_{2}$ processes $\frac{108}{653}$ unit;
  \item Load 2, second round: $P_{1}$ processes $\frac{464}{653}$ unit;
  \item Load 2, second round: $P_{2}$ processes $\frac{81}{653}$ unit.
  \end{compactitem}

  \begin{figure}
    \begin{center}
      {\includegraphics[width=0.75\linewidth]{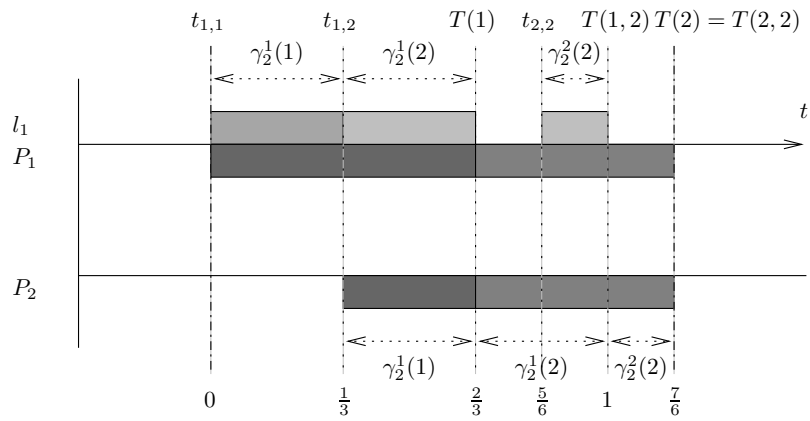}}
      \caption{The schedule of~\cite{WongVeBa05} for $\lambda=1$.\label{Fig:example_l1}}
    \end{center}
  \end{figure}

  This scheme gives us a total makespan equal to
  $\frac{781}{653}\frac{3}{4}\approx0.897$, which is (slightly) better
  than $0.9$. This shows that among the schedules having a total
  number of four installments, the solution of~\cite{WongVeBa05} is
  suboptimal.
\end{enumerate}

\subsection{Conclusion}

Despite its simplicity (two identical processors and two identical
loads), out motivating example clearly outlines the limitations of the
approach of~\cite{WongVeBa05}: this approach does not always return a
feasible solution and, when it does, this solution is not always
optimal. In the next section, we show how to compute an optimal
schedule when dividing each load into any prescribed number of
installments. Our simulations will later show that the gap between
\multiinst and the optimal schedule can be significantly large.

\section{Optimal solution}
\label{sec.new}

We now show how to compute an optimal schedule, when dividing each
load into any prescribed number of installments. Therefore, when this
number of installment is set to 1 for each load (i.e., $Q_n=1$, for
any $n$ in $[1,N]$), the following approach solves the problem
originally targeted by Min, Veeravalli, and Barlas.

To build our solution we use a linear programming approach. In fact,
we only have to list all the (linear) constraints that must be
fulfilled by a schedule, and write that we want to minimize the
$\mathrm{makespan}$. All these constraints are captured by the linear
program in Figure~\ref{linearprogram}. The optimality of the solution
comes from the fact that the constraints are exactly all the
constraints that any schedule must fulfill under the assumptions of
Section~\ref{sec.frame}, and a solution to the linear program is
obviously always feasible. This linear program simply encodes the
following constraints (a constraint has the same number below and in
Figure~\ref{linearprogram}):
\begin{compactenum}
\item $P_{i}$ cannot start a new communication to $P_{i+1}$ before the
  end of the corresponding communication from $P_{i-1}$ to $P_i$,

\item $P_{i}$ cannot start to receive the next installment of the
  $n$th load before having finished to send the current one to
  $P_{i+1}$,

\item $P_{i}$ cannot start to receive the first installment of the
  next load before having finished to send the last installment of the
  current load to $P_{i+1}$,

\item any transfer has to begin at a nonnegative time,

\item the duration of any transfer is equal to the product of the time
  taken to transmit a unit load by the volume of data to
  transfer,

\item processor $P_{i}$ cannot start to compute the $j$th installment
  of the $n$th load before having finished to receive the
  corresponding data,

\item the duration of any computation is equal to the product of the
  time taken to compute a unit load by the volume of the
  computation,

\item processor $P_i$ cannot start to compute the first installment of
  the next load before it has completed the computation of the last
  installment of the current load,

\item processor $P_i$ cannot start to compute the next installment of
  a load before it has completed the computation of the current
  installment of that load,

\item processor $P_i$ cannot start to compute the first installment of
  the first load before its availability date,

\item the portion of a load dedicated to a processor is necessarily
  nonnegative,
\item any load has to be completely processed,
\item the $\mathrm{makespan}$ is no smaller than the completion time
  of the last installment of the last load on any processor.
\end{compactenum}

\begin{figure*}
{\small
\begin{eqnarray}
\forall i < m-1, n\leq N,j \leq Q_{n}	\quad & \Ss{i+1,n,j} 		\quad & \geq \quad \Se{i,n,j} \label{cons1}\\
\forall i < m-1,n \leq N,j < Q_{n}	\quad & \Ss{i,n,j+1}		\quad & \geq \quad \Se{i+1,n,j} \label{cons2}\\
\forall i < m-1,n < N	 		\quad & \Ss{i,n+1,1}		\quad & \geq \quad \Se{i+1,n,Q_{n}} \label{cons3}\\
\forall i \leq m-1,n\leq N,j \leq Q_{n}	\quad & \Ss{i,n,j}		\quad & \geq \quad 0\label{cons4}\\
\forall i \leq m-1,n\leq N,j \leq Q_{n}	\quad & \Se{i,n,j} 		\quad & = \quad \Ss{i,n,j} + z_{i}V_{comm}(n)\sum_{k=i+1}^{m}\gamma_{k}^{j}(n)\label{cons5}\\
\forall i \geq 2, n\leq N, j\leq Q_{n}	\quad & \Cs{i,n,j}		\quad & \geq \quad \Se{i,n,j} \label{cons6}\\
\forall i\leq m,n\leq N,j\leq Q_{n}	\quad & \Ce{i,n,j}		\quad & = \quad \Cs{i,n,j} + w_{i}\gamma_{i}^{j}(n)V_{calc}(n)\label{cons7}\\
\forall i\leq m,n < N			\quad & \Cs{i,n+1,1}		\quad & \geq \quad \Ce{i,n,Q_{n}} \label{cons8}\\
\forall i\leq m,n \leq N,j < Q_{n}	\quad & \Cs{i,n,j+1}		\quad & \geq \quad \Ce{i,n,j} \label{cons9}\\
\forall i\leq m				\quad & \Cs{i,1,1}		\quad & \geq \quad \tau_{i} \label{cons10}\\
\forall i\leq m,n\leq N,j\leq Q_{n}	\quad & \gamma_{i}^{j}(n) 	\quad & \geq \quad 0\label{cons11}\\
\forall n\leq N				\quad & \sum_{i=1}^{m}\sum_{j=1}^{Q}\gamma_{i}^{j}(n) \quad & = \quad 1 \label{cons12}\\
\forall i \leq m			\quad & \mathrm{makespan}	\quad & \geq \quad \Ce{i,N,Q}\label{cons13}
\end{eqnarray}
}
\caption{The complete linear program.\label{linearprogram}}
\end{figure*}

Altogether, we have a linear program to be solved over the rationals,
hence a solution in polynomial time~\cite{karmarkar}. In practice,
standard packages like GLPK~\cite{glpk} will return the optimal
solution for all reasonable problem sizes.  Note that the linear
program gives the optimal solution for a prescribed number of
installments for each load. In the next section we discuss the problem
of the number of installments.

\section{Possible extensions}
\label{sec:ext}

Several of the model restrictions can be alleviated. First the model
uses \emph{uniform machines}, meaning that the speed of a processor
does not depend on the task that it executes. It is easy to extend the
linear program for unrelated parallel machines, introducing
$w_{i}^{n}$ to denote the time taken by $P_{i}$ to process a unit load
of type $n$. Also, all processors and loads are assumed to be
available from the beginning.  In our linear program, we have
introduced availability dates for processors. The same way, we could
have introduced release dates for loads.  Furthermore, instead of
minimizing the makespan, we could have targeted any other objective
function which is an affine combination of the loads completion time
and of the problem characteristics, like the average completion time,
the maximum or average (weighted) flow, etc.

The formulation of the problem does not allow any piece of the $n'$th
load to be processed before the $n$th load is completely processed, if
$n' > n$. We can easily extend our solution to allow for $N$ rounds of
the $N$ loads, each load being still divided into several
installments. This would allow to interleave the processing of the
different loads.

The divisible load model is linear, which causes major problems for
multi-installment approaches. Indeed, once we have a way to find an
optimal solution when the number of installments per load is given,
the question is: what is the optimal number of installments? Under a
linear model for communications and computations, the optimal number
of installments is infinite, as the following theorem states:

\begin{theorem}
  Assuming a linear cost model for communications and
  computations, consider any problem with one or more loads and
  at least two processors. Then, any schedule using a finite number of
  installments is suboptimal for makespan minimization.
\end{theorem}

This theorem is proved by building, from any schedule, another
schedule with a strictly smaller makespan. The proof is available in
the research report~\cite{GalletRoVi07a}.

An infinite number of installments obviously does not define a
feasible solution. Moreover, in practice, when the number of
installments becomes too large, the model is inaccurate, as acknowledged
in~\cite[pp. 224 and 276]{robertazzi96}. Any
communication incurs a startup cost $K$, which we express in bytes.
Consider the $n$th load, whose communication volume is $V_{comm}(n)$:
it is split into $Q_{n}$ installments, and each installment requires
$m-1$ communications.  The ratio between the actual and estimated
communication costs is roughly equal to
$\rho=\frac{(m-1)Q_{n}K+V_{comm}(n)}{V_{comm}(n)}>1$. Since $K$, $m$,
and $V_{comm}$ are known values, we can choose $Q_{n}$ such that
$\rho$ is kept relatively small, and so such that the model remains
valid for the target application. Another, and more accurate solution,
would be to introduce latencies in the model, as in~\cite{j89}. This latter
article shows how to design asymptotically optimal multi-installment
strategies for star networks. A similar approach could be used for
linear networks.

\section{Experiments}
\label{sec.experiments}

Using simulations, we now assess the relative performance of our linear
programming approach, of the solutions of~\cite{WongVe04,WongVeBa05},
and of simpler heuristics.  We first describe the experimental
protocol and then analyze the results.

\paragraph{Experimental protocol.}

We use Simgrid~\cite{legrand_ccgrid03} to simulate linear processor
networks.  Schedules are computed by a Perl script, and their validity
and theoretical makespan are checked before running them in the
simulator.

We study the following algorithms and heuristics:
\begin{compactitem}
\item The naive heuristic \naive distributes each load in a
  single installment and proportionally to the processor speeds.

\item The strategy for a single load, \singleload, presented by Min
  and Veeravalli in~\cite{WongVe04}. For each load, we set the time
  origin to the availability date of the first communication link (in
  order to try to prevent communication contentions).

\item The \singleinst strategy described in
  Section~\ref{sec.existingalgos}.

\item The \multiinst$n$ strategy. This is a slightly modified version
  of \multiinst which ensures that a load is not distributed in more
  than $n$ installments, the $n$th installment distributing all the
  load remaining work.

\item The \textsc{Heuristic B} presented by Min, Veeravalli and Barlas
  in~\cite{WongVeBa05}.

\item \LProg{n}: the solution of our linear program where each load is
  distributed in $n$ installments.
\end{compactitem}

We measure the relative performance of each heuristic on each
instance: we divide the makespan obtained by a given heuristic on a
given instance by the smallest makespan obtained, on that instance,
among all heuristics. Considering the relative performance enables us
to obtain meaningful statistics among instances with very different
makespans.

\paragraph{Instances.}

We emulate a heterogeneous linear network with $m=10$ processors.  We
consider two distribution types for processing powers:
\emph{homogeneous} where each processor $P_i$ has a processing power
$\frac{1}{w_{i}} = 100$ MFLOPS, and \emph{heterogeneous} where
processing powers are uniformly picked between $10$ and $100$ MFLOPS.
Communication link $l_i$ has a speed $\frac{1}{z_{i}}$ uniformly
chosen between $10$ Mb/s and $100$ Mb/s, and a latency between $0.1$
and $1$ ms (links with high bandwidths having small latencies).  For
homogeneous and heterogeneous platforms, simulation tasks have their
computation volumes either all uniformly distributed between $6$
GFLOPS and $4$ TFLOPS, or all uniformly distributed between $6$ and
$60$ GFLOPS.  For each combination of processing power distribution
and task size, we fix the communication to computation volume of all
tasks to either $0.01, 0.05, 0.1, 0.5, 1, 5, 10, 50$, or $100$
(bytes per FLOPS).  Each instance contains 50 loads. Finally, we
randomly built 100 instances per combination of the different
parameters, hence a total of 3,600 instances simulated and reported in
Table~\ref{table.overall}. The code and the experimental results can
be downloaded from:
\url{http://graal.ens-lyon.fr/~mgallet/downloads/DivisibleLoadsLinearNetwork.tar.gz}.

We fixed an upper-bound for the number of installments per load used
by the different heuristics: \multiinst to either $100$ or $300$,
\singleload to $100$, and \LProg{n} to either $1$, $2$, $3$, or $6$.

\paragraph{Discussions of the results.}

We first remark that the linear program approach always reaches the
best makespan. \LProg{1}, \LProg{2}, \LProg{3}, and \LProg{6} achieve
equivalent performance, always less than {0.5\%} away from
the optimal. This may seem counter-intuitive but can be readily
explained: multi-installment strategies mainly reduce the idle time
incurred on each processor before it starts processing the first task,
and the room for improvement is thus quite small in our (and~\cite{WongVeBa05}) batches of 50
tasks. The strict one-port communication model forbids the overlapping
of some communications due to different installments, and further
limits the room for performance enhancement.  Except in some peculiar
cases, distributing the loads in multi-installments do not induce
significant gains. In very special cases, \LProg{6} does not achieve
the best performance during the simulations, but this fact can be
explained by the latencies existing in simulations.

The bad performance of \naive, which can have makespans 8000 greater
than the optimal, justify the use of sophisticated scheduling
strategies. \singleinst has tremendously better performance than
\singleload as it far better takes into account communication link
availabilities: the huge difference of performance is due to the
instances with expensive communications. \singleinst achieves very
good average performance, within 6\% of the optimal. It also achieves
significantly better performance than \multiinst, and \heurB. This may
also be due to the fact that multi-installment strategies are not
efficient in our experimental context. The slight difference
performance between \multiinst 100 and \multiinst 300 shows that
\multiinst sometimes uses a large amount of installments for an
insignificant negative gain (certainly due to the latencies).  When
communication links are slow and when computations dominate
communications, \multiinst and \heurB can have makespans 98\% higher
than the optimal.


\begin{table}
		\centering
  \begin{tabular}{|l|r|l|r|}
	  \hline
	  \multicolumn{1}{|c|}{Heuristic} &  \multicolumn{1}{|c|}{Average} &  \multicolumn{1}{|c|}{Std dev.} &  \multicolumn{1}{|c|}{Max} \\
	  \hline
	  \!\naive &  \!\!1150.42887\!\! &  \!\!$1.6~10^{3}$\!\! &  8385.94163\!\! \\
	  \hline
	  \!\singleload 100\!\! &  \!\!1462.65842\!\! &  \!\!$2.0~10^{3}$\!\! &  \!\!10714.41753\!\! \\
	  \hline
	  \!\singleinst &  \!\!1.06307\!\! &  \!\!$8.0~10^{-2}$\!\! &  1.52324\!\! \\
	  \hline
	  \!\multiinst 100 &  \!\!1.13962\!\! &  \!\!$1.8~10^{-1}$\!\! &  1.98712\!\! \\
	  \hline
	  \!\multiinst 300 &  \!\!1.13963\!\! &  \!\!$1.8~10^{-1}$\!\! &  1.98712\!\! \\
	  \hline
	  \!\heurB &  \!\!1.13268\!\! &  \!\!$1.7~10^{-1}$\!\! &  2.01865\!\! \\
	  \hline
	  \!\LProg{1} &  \!\!1.00047\!\! &  \!\!$8.5~10^{-4}$\!\! &  1.00498\!\! \\
	  \hline
	  \!\LProg{2} &  \!\!1.00005\!\! &  \!\!$9.6~10^{-5}$\!\! &  1.00196\!\! \\
	  \hline
	  \!\LProg{3} &  \!\!1.00002\!\! &  \!\!$4.7~10^{-5}$\!\! &  1.00098\!\! \\
	  \hline
	  \!\LProg{6} &  \!\!1.00000\!\! &  \!\!0\!\! &  1.00001\!\! \\
	  \hline
  \end{tabular}
\caption{Summary of results.\label{table.overall}}
\end{table}



\section{Conclusion}
\label{sec:conclusion}

We have shown that a linear programming approach allows to solve all
instances of the scheduling problem addressed
in~\cite{WongVe04,WongVeBa05}. In contrast, the original approach was
providing a solution only for particular problem instances. Moreover,
the linear programming approach returns an optimal solution for any
number of installments, while the original approach was empirically
limited to very special strategies, and was often sub-optimal.

Intuitively, the solution of~\cite{WongVeBa05} is less efficient than
the schedule of Section~\ref{sec.ours} because it aims at locally
optimizing the makespan for the first load, and then optimizing the
makespan for the second one, and so on, instead of directly searching
for a global optimum. We were not able to provide closed-form
expressions characterizing optimal solutions, but, owing to the power
of linear programming, we were able to derive an optimal schedule for
any problem instance. We validated this approach through simulations
which confirmed that the best solution is always produced by the
linear programming approach, while solutions of~\cite{WongVeBa05} can
be far away from the optimal. The simulations also show that, in our
settings, the multi-installment strategies rarely lead to significant
gains.

\bibliographystyle{abbrv}
\bibliography{biblio}

\begin{thebibliography}{10}

\bibitem{AltilarPak98}
D.~Altilar and Y.~Paker.
\newblock An optimal scheduling algorithm for parallel video processing.
\newblock In {\em IEEE Int. Conference on Multimedia Computing and Systems},
  1998.

\bibitem{AltilarPak02}
D.~Altilar and Y.~Paker.
\newblock Optimal scheduling algorithms for communication constrained parallel
  processing.
\newblock In {\em Euro-Par 2002}, LNCS 2400, pages 197--206. Springer Verlag,
  2002.

\bibitem{j89}
O.~Beaumont, H.~Casanova, A.~Legrand, Y.~Robert, and Y.~Yang.
\newblock Scheduling divisible loads on star and tree networks: results and
  open problems.
\newblock {\em IEEE Trans. Parallel Distributed Systems}, 16(3):207--218, 2005.

\bibitem{robertazzi96}
V.~Bharadwaj, D.~Ghose, V.~Mani, and T.~Robertazzi.
\newblock {\em Scheduling Divisible Loads in Parallel and Distributed Systems}.
\newblock IEEE Computer Society Press, 1996.

\bibitem{BlazewiczDM99}
J.~Blazewicz, M.~Drozdowski, and M.~Markiewicz.
\newblock Divisible task scheduling - concept and verification.
\newblock {\em Parallel Computing}, 25:87--98, 1999.

\bibitem{ChanBG01}
S.~Chan, V.~Bharadwaj, and D.~Ghose.
\newblock Large matrix-vector products on distributed bus networks with
  communication delays using the divisible load paradigm: performance and
  simulation.
\newblock {\em Mathematics and Computers in Simulation}, 58:71--92, 2001.

\bibitem{GalletRoVi07a}
M.~Gallet, Y.~Robert, and F.~Vivien.
\newblock Comments on ``design and performance evaluation of load distribution
  strategies for multiple loads on heterogeneous linear daisy chain networks''.
\newblock Research report RR-6123, INRIA, 2007.
\newblock \url{http://hal.inria.fr/inria-00130294}.

\bibitem{GenaudGiVi2004}
S.~Genaud, A.~Giersch, and F.~Vivien.
\newblock Load-balancing scatter operations for grid computing.
\newblock {\em Parallel Computing}, 30(8):923--946, 2004.

\bibitem{cluster-special}
D.~Ghose and T.~Robertazzi, editors.
\newblock {\em Special issue on {\em Divisible Load Scheduling}}. Cluster
  Computing, 6, 1, 2003.

\bibitem{glpk}
{GLPK: GNU Linear Programming Kit}.
\newblock \url{http://www.gnu.org/software/glpk/}.

\bibitem{karmarkar}
N.~Karmarkar.
\newblock A new polynomial-time algorithm for linear programming.
\newblock In {\em {Proceedings of ACM STOC'84}}, pages 302--311, 1984.

\bibitem{LeeHam95}
C.~Lee and M.~Hamdi.
\newblock Parallel image processing applications on a network of workstations.
\newblock {\em Parallel Computing}, 21:137--160, 1995.

\bibitem{legrand_ccgrid03}
A.~Legrand, L.Marchal, and H.~Casanova.
\newblock {Scheduling Distributed Applications: The \textsc{SimGrid} Simulation
  Framework}.
\newblock In {\em Proceedings of CCGrid'03}, pages 138--145, May 2003.

\bibitem{LegrandSuVi06}
A.~Legrand, A.~Su, and F.~Vivien.
\newblock Minimizing the stretch when scheduling flows of biological requests.
\newblock In {\em Proceedings of SPAA '06}, pages 103--112. ACM Press, 2006.

\bibitem{LiBK03}
X.~Li, B.~Veeravalli, and C.~Ko.
\newblock Distributed image processing on a network of workstations.
\newblock {\em Int. J. Computers and Applications (ACTA Press)}, 25(2):1--10,
  2003.

\bibitem{Robertazzi-Computer2003}
T.~Robertazzi.
\newblock Ten reasons to use divisible load theory.
\newblock {\em IEEE Computer}, 36(5):63--68, 2003.

\bibitem{WangKMAC98}
R.~Wang, A.~Krishnamurthy, R.~Martin, T.~Anderson, and D.~Culler.
\newblock Modeling communication pipeline latency.
\newblock In {\em Measurement and Modeling of Computer Systems
  (SIGMETRICS'98)}, pages 22--32. ACM Press, 1998.

\bibitem{WongVe04}
H.~M. Wong and B.~Veeravalli.
\newblock Scheduling divisible loads on heterogeneous linear daisy chain
  networks with arbitrary processor release times.
\newblock {\em IEEE Trans. Parallel Distributed Systems}, 15(3):273--288, 2004.

\bibitem{WongVeBa05}
H.~M. Wong, B.~Veeravalli, and G.~Barlas.
\newblock Design and performance evaluation of load distribution strategies for
  multiple divisible loads on heterogeneous linear daisy chain networks.
\newblock {\em J. Parallel Distributed Computing}, 65(12):1558--1577, 2005.

\bibitem{LegrandNP}
Y.~Yang, H.~Casanova, M.~Drozdowski, M.~Lawenda, and A.~Legrand.
\newblock On the complexity of multi-round divisible load scheduling.
\newblock Research report RR-6096, INRIA, 2007.
\newblock \url{http://hal.inria.fr/inria-00123711}.

\end{thebibliography}
\end{document}